\documentclass[12pt]{article}
\title{
\vspace{-8mm}
\rightline{\small HUB--EP--98/73}
\vspace{-2mm}
\bf Confining Properties of\\ 
Abelian(-Projected) Theories}
\author{Dmitri Antonov 
\thanks{Telephone: 0049-30-2093 7974; Fax: 0049-30-2093 
7631; E-mail address: 
{\tt antonov@physik.hu-berlin.de}}{\,} 
\thanks {On leave of absence from the 
Institute of Theoretical and Experimental Physics (ITEP), 
B.Cheremushkinskaya 25, 
117 218 Moscow, Russia.}{\,} and Dietmar Ebert \thanks{E-mail address: 
{\tt debert@physik.hu-berlin.de}} 
\\
{\it Institut f\"ur Physik, Humboldt-Universit\"at zu Berlin,}\\
{\it Invalidenstra{\ss}e 110, D-10115 Berlin, Germany}}
\date{}
\begin{document}
\maketitle

\vspace{1mm}
\centerline{\bf {Abstract}}
\vspace{3mm}
\noindent
Representations of the Abelian-projected $SU(2)$- and 
$SU(3)$-gluodynamics in terms of the magnetic 
monopole currents are derived. Besides the quadratic part, 
the obtained effective actions contain interactions 
of these 
currents with the world-sheets of electric strings in 4D or electric 
vortex lines in 3D. 
Next, we illustrate that 3D compact QED is a small gauge boson 
mass limit of 3D Abelian Higgs model with external monopoles and  
give a physical interpretation to 
the confining string theory as the 
integral over the monopole densities. Finally, we derive the 
bilocal field strength correlator in the weak-field limit of 
3D compact QED, which turns out to be in line with the one predicted 
by the Stochastic Vacuum Model.

\newpage

\section{Introduction}

During several last years, there has appeared 
a vast amount of papers devoted to the 
description of confinement 
in Abelian-projected theories~\cite{th} (see 
e.g.~\cite{maedan, ball, kleinert, sha,  
orland, for, zubkov, over, mathur, euro, suzuki, komarov} 
and Refs. therein). The main goal of 
most of these papers is a derivation of the so-called string 
representation of such theories, i.e. a reformulation of their 
partition functions in terms of the integral over the world-sheets 
of the Abrikosov-Nielsen-Olesen (ANO) strings~\cite{ano} with a 
certain nonlocal (i.e. depending on a relative distance between 
two points 
in the target space) action. Such a 
representation then enables one to get the coupling constants 
of the corresponding string theory including higher order derivative 
terms 
and to evaluate correlators 
of the dual field strength tensors~\cite{euro, suzuki}, which play 
a major role in the so-called Stochastic Vacuum Model 
(SVM)~\cite{svm, ufn}. 
In the case when there are no external 
quarks in the underlying non-Abelian 
theory, 
the corresponding Abelian-projected theory is some kind 
of a dual Abelian Higgs model with magnetic Higgs fields, which 
describe the  
condensates of monopole Cooper pairs. This model possesses 
classical solutions, which in 4D are just the 
electric ANO strings, whereas in 3D they are 
simply closed electric vortex lines~\cite{for}. It is therefore 
intuitively clear that there should exist some interaction between 
magnetic monopoles and ANO strings (vortex lines), and 
we shall demonstrate below that such an interaction really exists. 
Besides the interaction of the monopole Cooper pairs   
with strings or vortex lines, we 
shall obtain the full effective action of the monopole currents 
for the $SU(2)$- and $SU(3)$ Abelian-projected theories. Notice, 
that an action of this kind for the case of the usual AHM on a 
lattice has been presented in Ref.~\cite{over} in a form where 
the above 
mentioned interaction of the monopole currents with the string 
world-sheets has not, however, been manifestly shown. 

Another, known for even a 
longer time theory, which allows for an analytic 
description of confinement, is 3D compact QED~\cite{for, 
mon}. There, similarly to the Abelian-projected theories, 
confinement also occurs due to the monopole condensation. 
The problem of string representation 
of this model has 
been addressed in Ref.~\cite{kleinert}, where it has been argued 
that the desired string theory is formulated in terms of a massive 
antisymmetric tensor field (usually called Kalb-Ramond field~\cite{kalb}) 
interacting with the string world-sheet. The complete form of 
the action of this field 
turned out to be quite a nonlinear one and was found 
recently in Ref.~\cite{confstr}. After that, the resulting theory, which 
is usually referred to as the confining string theory, has undergone 
intensive developments~\cite{diamant}. 
Notice, that  
the weak-field effective action of this theory occurs to have  
the linear form of the massive Kalb-Ramond field action, coinciding 
with the one of the dual version of AHM in the London limit. 
This reflects the fact that there should exist a correspondence between 
the direct formulations of compact QED and AHM as well. Below, we shall 
illustrate this to be really the case, namely that 
3D compact QED is nothing else, but the small gauge boson mass 
limit of 3D AHM with external monopoles. Besides that, we show 
that the confining string theory is actually simply the 
integral over the monopole densities, which gives physical 
interpretation to the Kalb-Ramond field as a sum of the monopole 
and photon field strength tensors. 
Finally, we evaluate field correlators 
in the weak-field limit of 3D compact QED, and argue that their 
large-distance asymptotic behaviours are  
in line with the ones observed in the lattice experiments~\cite{digiac} 
within SVM.   

The organization of the paper is as follows. In the next Section, we 
study the topic of representation of Abelian-projected theories 
in terms of the magnetic monopole currents. In 
Section 3, we revisit 3D compact QED and its string representation, 
after which it is demonstrated 
how this theory can be obtained by a limiting 
procedure from 3D AHM with monopoles.   
In the Appendix, we outline some details of the path-integral duality 
transformation.

\section{Representation of the Abelian-Projected Theories in Terms 
of Monopole Currents}

Let us start with the 4D $SU(2)$ Abelian-projected gluodynamics, which 
is argued to be just DAHM, whose partition function in the London limit 
has the form (see e.g.~\cite{euro})   

\begin{equation}
\label{odin}
{\cal Z}_{\rm 4D{\,}DAHM}=\int DB_\mu D\theta^{{\rm sing.}} 
D\theta^{{\rm reg.}}\exp\left\{-\int d^4x\left[\frac14
F_{\mu\nu}^2+\frac{\eta^2}{2}
\left(\partial_\mu\theta-2gB_\mu\right)^2\right]\right\}. 
\end{equation}
Here, $F_{\mu\nu}=\partial_\mu B_\nu-\partial_\nu B_\mu$ is a 
field strength tensor of the dual vector potential $B_\mu$, and $g$ is a 
magnetic coupling constant ($2g$ is the magnetic charge of the 
monopole Cooper pair). Next, 
$\eta$ denotes the v.e.v. of the magnetic Higgs field, 
whose phase has the form 
$\theta=\theta^{\rm sing.}+\theta^{\rm reg.}$, 
where $\theta^{\rm sing.}$ describes a given electric string 
configuration, whereas $\theta^{\rm reg.}$ stands for a single-valued 
fluctuation around this configuration. 
The singular part of the phase of the 
magnetic Higgs field is related to the (closed) world-sheet 
$\Sigma$ of the electric ANO string according to the equation 

\begin{equation}
\label{dva}
\varepsilon_{\mu\nu\lambda\rho}\partial_\lambda
\partial_\rho\theta^{{\rm sing.}}(x)=2\pi\Sigma_{\mu\nu}(x) 
\equiv 2\pi\int\limits_{\Sigma}^{}d\sigma_{\mu\nu}(x(\xi))
\delta(x-x(\xi))
\end{equation}
where $\Sigma_{\mu\nu}$ is usually referred to as a vorticity tensor 
current~\cite{orland}, and 
$\xi=\left(\xi^1,\xi^2\right)$ stands for the two-dimensional 
coordinate. Performing the path-integral duality 
transformation~\cite{orland, for, euro}, we arrive at the following 
representation for the partition function~(\ref{odin}) (see Appendix A)

\begin{equation}
\label{mon1}
{\cal Z}_{\rm 4D{\,}DAHM}=\int DA_\mu Dx_\mu(\xi) Dh_{\mu\nu}\exp\Biggl\{
-\int d^4x\Biggl[\frac{1}{12\eta^2}H_{\mu\nu\lambda}^2-i\pi
h_{\mu\nu}\Sigma_{\mu\nu}+\left(gh_{\mu\nu}+\partial_\mu A_\nu-
\partial_\nu A_\mu\right)^2\Biggr]\Biggr\}.
\end{equation}
Here, $A_\mu$ is the usual gauge field dual to the vector potential 
$B_\mu$, and 
$H_{\mu\nu\lambda}\equiv\partial_\mu h_{\nu\lambda}+
\partial_\lambda h_{\mu\nu}+\partial_\nu h_{\lambda\mu}$ is the field 
strength 
tensor of a massive antisymmetric tensor field $h_{\mu\nu}$ (the so-called 
Kalb-Ramond field~\cite{kalb}). This antisymmetric spin-1 tensor field 
describes a  
massive dual vector boson. Thus, the path-integral duality transformation 
is just a way of getting a coupling of this boson to a string world-sheet, 
rather than to a world-line (as it takes place in the usual 
case of the Wilson loop). In particular, carrying out in Eq.~(\ref{mon1}) 
the Gaussian integration over the Kalb-Ramond field, one gets 
a representation of the partition function in terms of an interaction 
of the elements of the world-sheet $\Sigma$, mediated by the 
propagator of this field~\cite{euro}. In what follows, we shall 
derive another useful representation for the partition function, 
where it will be expressed directly in terms of magnetic monopole 
currents. 

Notice that according to the equation of motion for the field $A_\mu$,  
the absence of external electric currents is expressed by the 
equation $\partial_\mu {\cal F}_{\mu\nu}=0$, where 
${\cal F}_{\mu\nu}\equiv \partial_\mu A_\nu-\partial_\nu A_\mu+
gh_{\mu\nu}$. 
Regarding ${\cal F}_{\mu\nu}$ 
as a full electromagnetic field strength tensor, one can write for it 
the corresponding Bianchi identity modified by the monopoles, 
$\partial_\mu\tilde {\cal F}_{\mu\nu}=
g\partial_\mu\tilde h_{\mu\nu}$. This identity means that the 
monopole 
current can be written in terms of the Kalb-Ramond field $h_{\mu\nu}$ 
as 

\begin{equation}
\label{jmon}
j_\mu=g\partial_\nu\tilde h_{\nu\mu},
\end{equation}
which manifests its 
conservation.

It is also instructive to write down the equation of motion for the 
Kalb-Ramond field in terms of the introduced full electromagnetic 
field strength tensor. This equation has the form 
${\cal F}_{\nu\lambda}=\frac{g}{m^2}\partial_\mu H_{\mu\nu\lambda}+
\frac{i\pi}{2g} \Sigma_{\nu\lambda}$, where  
$m=2g\eta$ stands for 
the mass of the dual gauge boson (equal to the 
mass of the Kalb-Ramond field). 
By virtue of conservation of the vorticity tensor current for the 
closed string world-sheets, $\partial_\mu
\Sigma_{\mu\nu}=0$, this equation again yields the condition of 
absence of external electric currents, $\partial_\mu   
{\cal F}_{\mu\nu}=0$.

Let us now turn ourselves to a derivation of the 
monopole current representation for the partition function of DAHM. 
To this end, we shall first resolve the equation $\frac{g}{2}
\varepsilon_{\mu\nu\lambda\rho}\partial_\nu h_{\lambda\rho}=-j_\mu$ 
w.r.t. $h_{\mu\nu}$, which yields 

$$h_{\mu\nu}(x)=-\frac{1}{2\pi^2g^2}\varepsilon_{\mu\nu\lambda\rho}
\int d^4y\frac{(x-y)_\lambda}{|x-y|^4}j_\rho (y).$$
Next, we get the following expressions for various terms 
on the R.H.S. of Eq.~(\ref{mon1})

$$H_{\mu\nu\lambda}^2=\frac{6}{g^2}j_\mu^2,~~  
\int d^4xh_{\mu\nu}^2=\frac{1}{2\pi^2g^2}\int d^4xd^4yj_\mu(x)
\frac{1}{(x-y)^2}j_\mu(y).$$
Bringing all this together 
and performing in Eq.~(\ref{mon1}) 
the hypergauge transformation 
$h_{\mu\nu}\to h_{\mu\nu}+\partial_\mu\Lambda_\nu-
\partial_\nu\Lambda_\mu$ with the gauge function $\Lambda_\mu=
-\frac{1}{g}A_\mu$, which eliminates 
the field $A_\mu$, 
we finally arrive at the desired 
monopole current representation, which has the form 

$$
{\cal Z}_{\rm 4D{\,}DAHM}=
\int Dx_\mu(\xi) Dh_{\mu\nu}\exp\Biggl\{-\frac{1}{2\pi^2}
\int d^4xd^4yj_\mu(x)\frac{1}{(x-y)^2}j_\mu(y)-\frac{2}{m^2}
\int d^4xj_\mu^2+
$$

\begin{equation}
\label{mon2}
+\frac{2\pi i}{g}S_{\rm int.}(\Sigma, j_\mu)
\Biggr\}.
\end{equation}
The first term in the exponent on the 
R.H.S. of Eq.~(\ref{mon2}) 
has the form of the Biot-Savart energy of the electric field generated 
by monopole currents~\cite{for}, the second term corresponds to 
the (gauged) kinetic energy of Cooper pairs,   
and the term 

\begin{equation}
\label{sint}
S_{\rm int.}(\Sigma, j_\mu)=
\frac{1}{4\pi^2}\varepsilon_{\mu\nu\lambda\rho}
\int d^4xd^4y j_\mu(x)\frac{(y-x)_\nu}{|y-x|^4}\Sigma_{\lambda\rho}(y)
\end{equation}
describes the 
interaction of the string world-sheet with the monopole 
current $j_\mu$. This interaction can obviously be rewritten 
in the form $S_{\rm int.}=\int d^4x j_\mu H_\mu^{\rm str.}$, where 
$H_\mu^{\rm str.}$ is the four-dimensional analogue of the 
magnetic induction, produced by the 
electric string according to the equation

\begin{equation}
\label{induct}
\varepsilon_{\mu\nu\lambda\rho}\partial_\lambda H_\rho^{\rm str.}=
\Sigma_{\mu\nu}.
\end{equation}  
Notice, that if one includes an additional 
current describing an external monopole,

\begin{equation}
\label{part}
j_\mu^{\rm ext.}(x)=g\oint\limits_{\Gamma}^{}dx_\mu(\tau)
\delta(x-x(\tau)),
\end{equation} 
there arises among others an 
interaction term~(\ref{sint}), which in this case takes the form 
$S_{\rm int.}=g\hat L(\Sigma, \Gamma)$, 
where $\hat L(\Sigma, 
\Gamma)$ is simply the 
Gauss
linking number of the world-sheet $\Sigma$ 
with the contour $\Gamma$~\footnote{Topological 
interactions of this kind are sometimes interpreted as a 4D analogue 
of the Aharonov-Bohm effect. In particular, this interaction, albeit 
for the current of an external electrically charged 
particle with the string 
world-sheet, emerges in the string representation for the 
Wilson loop of this particle in AHM~\cite{zubkov}.}. 

Clearly, the functional integral over the Kalb-Ramond 
field in Eq.~(\ref{mon2}) has to be evaluated at the saddle-point 

$$
h_{\mu\nu}^{\rm s.p.}(x)=\frac{ig\eta^3}{\pi}\int\limits_{\Sigma}^{}
d\sigma_{\mu\nu}(x(\xi))\frac{K_1(m|x-x(\xi)|)}{|x-x(\xi)|},
$$
where from now on $K_n$'s, $n=0,1,2$, stand for the modified Bessel 
functions. 
By virtue of Eq.~(\ref{jmon}), 
the monopole current can then be expressed via the string world-sheet 
$\Sigma$ as follows

$$
j_\mu(x)=\frac{im^2\eta}{8\pi}\varepsilon_{\mu\nu\lambda\rho}
\int\limits_{\Sigma}^{}d\sigma_{\lambda\rho}(x(\xi))
\frac{(x-x(\xi))_\nu}{(x-x(\xi))^2}\cdot
$$

$$
\cdot\Biggl\{
\frac{K_1(m|x-x(\xi)|)}{|x-x(\xi)|}+\frac{m}{2}\Biggl[
K_0(m|x-x(\xi)|)+K_2(m|x-x(\xi)|)\Biggr]\Biggr\}.
$$

It is straightforward to extend the above analysis to the case 
of the Abelian-projected $SU(3)$-gluodynamics~\cite{maedan}. 
There, the partition function~(\ref{odin}) is replaced by 

\begin{equation}
\label{suz2}
{\cal Z}_{SU(3)}=\int D\vec B_\mu D\theta_a^{\rm sing.}
D\theta_a^{\rm reg.} \delta\left(\sum\limits_{a=1}^{3}
\theta_a\right)
\exp\Biggl\{-\int d^4x\Biggl[
\frac14\vec F_{\mu\nu}^2+\frac{\eta^2}{2}\sum\limits_{a=1}^{3}
\left(\partial_\mu\theta_a-g\vec\varepsilon_a\vec B_\mu\right)^2
\Biggr]\Biggr\} 
\end{equation}
with 
$\vec F_{\mu\nu}=\partial_\mu \vec B_\nu-\partial_\nu 
\vec B_\mu$ standing for the field strength tensor of the Abelian 
vector potential $\vec B_\mu\equiv\left(B_\mu^3,B_\mu^8\right)$  
dual to the usual vector potential    
$\vec A\equiv\left(A_\mu^3,A_\mu^8\right)$. 
Next, on the R.H.S. of Eq.~(\ref{suz2}), 

$$\vec \varepsilon_1=\left(1,0\right),~ \vec\varepsilon_2=\left(-\frac12,
-\frac{\sqrt{3}}{2}\right),~ \vec\varepsilon_3=\left(-\frac12, 
\frac{\sqrt{3}}{2}\right)$$
denote the so-called root vectors, and the phases of three magnetic 
Higgs fields, $\theta_a$'s, obey the constraint 
$\sum\limits_{a=1}^{3}
\theta_a=0$, 
which is due to the fact that the unitary group under 
study is special. The singular parts of these fields are related 
to the world-sheets of the strings of three types as

\begin{equation}
\label{suz3}
\varepsilon_{\mu\nu\lambda\rho}\partial_\lambda\partial_\rho
\theta_a^{\rm sing.}(x)=2\pi\Sigma_{\mu\nu}^a(x)\equiv 
2\pi\int\limits_{\Sigma_a}^{}d\sigma_{\mu\nu}(x_a(\xi))
\delta(x-x_a(\xi)),
\end{equation}
where $x_a\equiv x_\mu^a(\xi)$ is a four-vector parametrizing 
the world-sheet $\Sigma_a$. 

Performing the path-integral duality transformation of Eq.~(\ref{suz2}) 
by making use 
of Eq.~(\ref{suz3}), we obtain (see Ref.~\cite{suzuki} for the 
details)

$$
{\cal Z}_{SU(3)}=\int Dx_\mu^a(\xi)\delta\left(\sum\limits_{a=1}^{3}
\Sigma_{\mu\nu}^a\right)DA_\mu^aDh_{\mu\nu}^a\exp\left\{
-\int d^4x\left[\frac{1}{12\eta^2}\left(H_{\mu\nu\lambda}^a
\right)^2-\right.\right.
$$

\begin{equation}
\label{su3}
\left.\left.
-i\pi h_{\mu\nu}^a\Sigma_{\mu\nu}^a+\left(g\frac{\sqrt{3}}{2\sqrt{2}}
h_{\mu\nu}^a+\partial_\mu A_\nu^a-\partial_\nu A_\mu^a\right)^2
\right]\right\},
\end{equation}
where $A_\mu^a\equiv\vec \varepsilon_a\vec A_\mu$. Eq.~(\ref{su3}) 
means that the three monopole currents can be expressed in terms of 
three Kalb-Ramond fields as $j_\mu^a=
g\frac{\sqrt{3}}{2\sqrt{2}}
\partial_\nu \tilde h_{\nu\mu}^a$ ({\it cf.} Eq.~(\ref{jmon})). 
Finally, rewriting Eq.~(\ref{su3}) via these currents and 
resolving the constraint $\sum\limits_{a=1}^{3}\Sigma_{\mu\nu}^a=0$ 
by integrating over one of the world-sheets (for concreteness, 
$x_\mu^3(\xi)$), we obtain 

$$
{\cal Z}_{SU(3)}=\int Dx_\mu^1(\xi)Dx_\mu^2(\xi)
Dh_{\mu\nu}^a
\exp\Biggl\{-\frac{1}{2\pi^2}
\int d^4xd^4yj_\mu^a(x)\frac{1}{(x-y)^2}j_\mu^a(y)-\frac{2}{m_B^2}
\int d^4x\left(j_\mu^a\right)^2+
$$

\begin{equation}
\label{su31}
+4\pi i\sqrt{
\frac23}\frac{1}{g}\Biggl[S_{\rm int.}\left(\Sigma^1, j_\mu^1\right)+
S_{\rm int.}\left(\Sigma^2, j_\mu^2\right)-S_{\rm int.}
\left(\Sigma^1, j_\mu^3\right)-S_{\rm int.}
\left(\Sigma^2, j_\mu^3\right)\Biggr]\Biggr\},
\end{equation}
where $m_B=\sqrt{\frac{3}{2}}
g\eta$ stands for the mass of the fields $B_\mu^3$ and $B_\mu^8$, 
which they acquire due to the Higgs mechanism. Eq.~(\ref{su31}) 
is the desired representation for the partition function of the 
Abelian-projected $SU(3)$-gluodynamics in terms of three monopole 
currents, which should be evaluated at the saddle-point. 
The terms in square brackets on its R.H.S. yield an interference between 
various possibilities of the interaction between the string 
world-sheets and monopole currents in this model 
to occur.

For illustrations, 
let us establish a correspondence 
of the above 
results to the 3D ones. Namely, let us derive 
a 3D analogue of 
Eq.~(\ref{mon2}), i.e. find a representation in terms of the 
monopole currents of the dual Ginzburg-Landau model. There, 
Eq.~(\ref{dva}) is replaced by~\cite{for}

\begin{equation}
\label{vortex}
\varepsilon_{\mu\nu\lambda}\partial_\nu\partial_\lambda
\theta^{\rm sing.}\left(\vec x{\,}\right)=2\pi\delta_\mu
\left(\vec x{\,}\right).
\end{equation}
Here, on the R.H.S. stands the so-called vortex density with 
$\delta_\mu\left(\vec x{\,}\right)\equiv\int\limits_{L}^{}
dy_\mu(\tau)\delta\left(\vec x-\vec y(\tau)\right)$ being the 
transverse $\delta$-function defined w.r.t. the electric 
vortex line $L$, parametrized by the vector 
$\vec y(\tau)$. This line is closed in the case under study, 
i.e. in the absence of external quarks, which means that $\partial_\mu 
\delta_\mu=0$. Performing again by virtue of Eq.~(\ref{vortex}) 
the path-integral duality 
transformation of the partition function~(\ref{odin}) with the 3D action, 
we get for it the 
following representation

$$
{\cal Z}_{\rm 3D{\,}DAHM}=\int D\varphi Dy_\mu(\tau)Dh_\mu
\exp\Biggl\{-\int d^3x\Biggl[\frac{1}{4\eta^2}\left(\partial_\mu
h_\nu-\partial_\nu h_\mu\right)^2-
$$

\begin{equation}
\label{threed}
-2\pi ih_\mu\delta_\mu
+\left(g\sqrt{2}h_\mu+\partial_\mu\varphi\right)^2\Biggr]
\Biggr\}.
\end{equation}
Notice, that the Kalb-Ramond field has now reduced to a massive one-form 
field $h_\mu$ with the mass $m=2g\eta$, 
as well as the $A_\mu$-field reduced to a 
scalar $\varphi$. Analogously to the 4D case, the field 
${\cal E}_\mu\equiv g\sqrt{2}h_\mu+\partial_\mu\varphi$ 
can be regarded as a full electric field, defined via the 
full dual electromagnetic field strength tensor as ${\cal E}_\mu=
\frac12\varepsilon_{\mu\nu\lambda}{\cal F}_{\nu\lambda}$. 
The absence of external quarks is now expressed by the equation 
$\partial_\mu {\cal E}_\mu=0$, following from the equation of motion 
for the field $\varphi$. 
Correspondingly, the monopole currents are defined as 
$j_\nu=\partial_\mu {\cal F}_{\mu\nu}=g\sqrt{2} 
\varepsilon_{\mu\nu\lambda}\partial_\mu h_\lambda$ and are 
manifestly conserved. Notice also, that the condition 
of closeness of the vortex lines,  
$\partial_\mu\delta_\mu=0$, unambiguously exhibits itself as a condition 
of absence of external quarks, $\partial_\mu {\cal E}_\mu=0$, by virtue 
of equation of motion for the field $h_\mu$, which can be 
written in the form 
${\cal E}_\mu=\frac{1}{g\sqrt{2}}\left[\frac{1}{2\eta^2}
\partial_\nu\left(\partial_\nu h_\mu-\partial_\mu h_\nu\right)+
i\pi\delta_\mu\right]$.
 
Next, after performing 
the gauge transformation $h_\mu\to h_\mu+\partial_\mu 
\gamma$ with the gauge function $\gamma=-\frac{1}{g\sqrt{2}}\varphi$, 
the field $\varphi$ drops out. 
Expressing $h_\mu$ 
via $j_\mu$ as follows 

$$
h_\mu\left(\vec x{\,}\right)=-\frac{1}{4\sqrt{2}\pi g}
\varepsilon_{\mu\nu\lambda}\frac{\partial}{\partial x_\nu}
\int d^3y\frac{j_\lambda\left(\vec y{\,}\right)}{\left|
\vec x-\vec y{\,}\right|}
$$
and substituting this expression into the R.H.S. of Eq.~(\ref{threed}), 
we finally arrive at the desired representation for the partition 
function of 3D DAHM in terms of the monopole currents

$$
{\cal Z}_{\rm 3D{\,}DAHM}=\int Dy_\mu(\tau)Dh_\mu\exp\Biggl\{
-\Biggl[\frac{1}{4\pi}\int d^3xd^3yj_\mu\left(\vec x{\,}\right)
\frac{1}{\left|\vec x-\vec y{\,}\right|}j_\mu\left(\vec y{\,}\right)+
$$

\begin{equation}
\label{threed1}
+\frac{1}{m^2}\int d^3xj_\mu^2+\frac{\sqrt{2}\pi i}{g}
S_{\rm int.}(L, j_\mu)\Biggr]\Biggr\}.
\end{equation}
The 
interaction term of the electric vortex line with the 
monopole current now takes the form

$$
S_{\rm int.}(L, j_\mu)=\frac{1}{4\pi}\varepsilon_{\mu\nu\lambda}
\int d^3xd^3yj_\mu\left(\vec x{\,}\right)\frac{\left(\vec y-\vec x{\,}
\right)_\nu}{\left|\vec y-\vec x{\,}\right|^3}\delta_\lambda
\left(\vec y{\,}\right).
$$
This interaction 
term can be again rewritten as $S_{\rm int.}=\int d^3x j_\mu 
H_\mu^{\rm vor.}$, where the magnetic induction, generated by 
the electric vortex  
line, obeys the equation $\varepsilon_{\mu\nu\lambda}
\partial_\nu H_\lambda^{\rm vor.}=\delta_\mu$.  
In the particular case, when one introduces an external 
current of 
the form~(\ref{part}), there emerges a term 
$S_{\rm int.}=g
\hat L(L, \Gamma)$ with $\hat L(L, \Gamma)$ 
standing for the Gauss linking number 
of the contours $L$ and $\Gamma$. The functional integral 
over the field $h_\mu$ in Eq.~(\ref{threed1}) should again be evaluated 
at the saddle-point $h_\mu^{\rm s.p.}$, which is determined by 
the classical equation of motion, following from Eq.~(\ref{threed}) 
after gauging away the field $\varphi$. This saddle-point has the 
form 

$$
h_\mu^{\rm s.p.}
\left(\vec x{\,}\right)=\frac{i\eta^2}{2}\oint\limits_{L}^{}
dy_\mu(\tau)\frac{{\rm e}^{-m\left|\vec x-\vec y(\tau)\right|}}{
\left|\vec x-\vec y(\tau)\right|},
$$
which yields the following expression for the monopole current 

$$
j_\mu\left(\vec x{\,}\right)=\frac{ig\eta^2}{\sqrt{2}}
\varepsilon_{\mu\nu\lambda}\oint\limits_{L}^{}dy_\lambda(\tau)
\frac{\left(\vec x-\vec y(\tau)\right)_\nu}{
\left(\vec x-\vec y(\tau)\right)^2}\left(m+\frac{1}{\left|\vec x-
\vec y(\tau)\right|}\right){\rm e}^{-m\left|\vec x-\vec y(\tau)\right|}.
$$

In the next Section, we shall investigate the relation between 
3D AHM with external monopoles and 3D compact QED.

\section{Vacuum Correlators and String Representation of 3D 
Compact QED}

In this Section, we shall revisit 
3D compact QED and find its 
string representation in the form of an integral over 
the monopole densities. Besides that, we shall investigate 
vacuum correlators 
in the weak-field limit, and demonstrate the relation of this theory to 
3D AHM with monopoles.

The most important feature of 3D {\it compact} QED, which 
distinguishes it from the noncompact case, is 
the existence of magnetic 
monopoles. Their general configuration is the Coulomb gas with the 
action~\cite{mon}

\begin{equation}
\label{smon}
S_{\rm mon.}=g^2\sum\limits_{a<b}^{}q_aq_b\left(\Delta^{-1}
\right)\left(\vec z_a, \vec z_b\right)+S_0\sum\limits_{a}^{}
q_a^2,
\end{equation}
where $\Delta$ is the 3D Laplace operator, and 
$S_0$ is the action of a single monopole, $S_0=\frac{{\rm const.}}{e^2}$.
Here, similarly to Ref.~\cite{mon}, we have adopted  
standard Dirac notations, where $eg=2\pi n$, 
restricting ourselves to the monopoles of the minimal charge, i.e. 
setting $n=1$. Then, the partition function of the grand canonical 
ensemble of monopoles associated with the 
action~(\ref{smon}) reads 

\begin{equation}
\label{zmon}
{\cal Z}_{\rm mon.}=\sum\limits_{N=1}^{+\infty}\sum\limits_{q_a=\pm 1}^{}
\frac{\zeta^N}{N!}\prod\limits_{i=1}^{N}\int d^3z_i
\exp\left[-\frac{\pi}{2e^2}\int d^3xd^3y \rho_{\rm gas}\left(\vec x{\,}
\right)\frac{1}{\left|\vec x-\vec y{\,}\right|}\rho_{\rm gas}\left(
\vec y{\,}\right)\right],
\end{equation}
where $\rho_{\rm gas}\left(\vec x{\,}\right)=\sum\limits_{a}^{}q_a
\delta\left(\vec x-\vec z_a\right)$ is the monopole density, 
corresponding to the gas configuration. 
Here, a single monopole weight $\zeta\propto\exp\left(-S_0
\sum\limits_a^{}q_a^2\right)$ has the dimension of ${\rm mass}^3$ 
and is usually referred to as fugacity. Notice also that, as usual,  
we have 
restricted ourselves to the values $q_a=\pm 1$, since 
at large 
values of the magnetic coupling constant $g$, monopoles 
with $|q|>1$ turn out to be unstable and 
tend to dissociate into the monopoles with $|q|=1$.
Below in this Section, it will be demonstrated 
that the limit of a small gauge boson mass (which takes place e.g. at 
large $g$) is just 
the case, when 3D compact QED follows from 3D AHM 
with external monopoles. 

Next, Coulomb interaction can be made local, albeit nonlinear one, 
by introduction of an 
auxiliary scalar field~\cite{mon}

\begin{equation}
\label{zcos}
{\cal Z}_{\rm mon.}=\int D\chi\exp\left\{-\int d^3x
\left[\frac12\left(\partial_\mu\chi\right)^2-2\zeta
\cos (g\chi)\right]\right\}.
\end{equation}
The magnetic mass $m=g\sqrt{2\zeta}$ of the field $\chi$, 
following from the quadratic term in the expansion of the 
cosine on the R.H.S. of Eq.~(\ref{zcos}), is due to the Debye 
screening in the monopole plasma. The next, quartic, term of the 
expansion determines the coupling constant of the 
diagrammatic expansion for the monopole gas, which is therefore 
exponentially small and proportional to $g^4\exp\left(-{\rm const.}
g^2\right)$. 

Let us now cast the partition function~(\ref{zcos}) into the form 
of an integral over the monopole densities. This can be done by 
introducing into Eq.~(\ref{zmon}) a unity of the form 

$$
\int D\rho\delta\left(\rho\left(\vec x{\,}\right)-\rho_{\rm gas}\left(
\vec x{\,}\right)
\right)=\int D\rho D\mu\exp\left\{i\left[\int d^3x\mu\rho-
\sum\limits_{a}^{}q_a\mu\left(\vec z_a\right)\right]\right\}.
$$
Then, passing to the representation of the partition function in 
terms of the field $\chi$, 
changing the integration variable, $\mu\to\phi\equiv
g\chi-\mu$, and integrating the field $\chi$ out, we get

\begin{equation}
\label{zrhophi}
{\cal Z}_{\rm mon.}=
\int D\rho D\phi\exp\left\{-\frac{\pi}{2e^2}\int d^3xd^3y
\rho\left(\vec x{\,}\right)\frac{1}{\left|\vec x-\vec y{\,}\right|}
\rho\left(\vec y{\,}\right)+\int d^3x\left(2\zeta\cos\phi-i\phi\rho
\right)\right\}.
\end{equation}
Finally, integrating over the field $\phi$ by resolving the corresponding 
saddle-point equation, 

\begin{equation}
\label{speq}
\sin\phi=-\frac{i\rho}{2\zeta},
\end{equation}
we arrive at the desired representation for the 
partition function 

\begin{equation}
\label{zdens}
{\cal Z}_{\rm mon.}=
\int D\rho \exp\left\{-\left[\frac{\pi}{2e^2}\int d^3xd^3y
\rho\left(\vec x{\,}\right)\frac{1}{\left|\vec x-\vec y{\,}\right|}
\rho\left(\vec y{\,}\right)+V[\rho]\right]\right\},
\end{equation}
where 

\begin{equation}
\label{pot}
V[\rho]=\int d^3x\left\{\rho\ln\left[\frac{\rho}{2\zeta}+
\sqrt{1+\left(\frac{\rho}{2\zeta}\right)^2}\right]-2\zeta
\sqrt{1+\left(\frac{\rho}{2\zeta}\right)^2}\right\}
\end{equation}
is the parabolic-type effective monopole potential, whose asymptotic 
behaviours at $\rho\ll\zeta$ and $\rho\gg\zeta$ read 

\begin{equation}
\label{lowdens}
V[\rho]\longrightarrow\int d^3x\left(-2\zeta+\frac{\rho^2}{4\zeta}
\right)
\end{equation}
and 

$$
V[\rho]\longrightarrow\int d^3x\left[\rho\left(\ln\frac{\rho}{\zeta}
-1\right)\right],
$$
respectively. 
Notice, that during the integration over the field $\phi$ 
in Eq.~(\ref{zrhophi}),  
we have chosen only the real branch of the solution to the 
saddle-point equation~(\ref{speq}) and 
disregarded the complex ones. 

The obtained representation for the partition function in terms of the 
monopole densities can be immediately applied to the calculation 
of the coefficient function ${\cal D}^{\rm mon.}\left(x^2\right)$, 
related to the bilocal correlator of the field strength tensors 
as follows~\cite{svm, ufn}

$$\left<{\cal F}_{\lambda\nu}\left(\vec x{\,}\right) 
{\cal F}_{\mu\rho}(0)
\right>_{A_\mu,\rho}=
\Biggl(\delta_{\lambda\mu}\delta_{\nu\rho}-\delta_{\lambda\rho}
\delta_{\nu\mu}\Biggr){\cal D}^{\rm mon.}\left(x^2\right)+$$

\begin{equation}
\label{cora}
+\frac12\Biggl[\partial_\lambda
\Biggl(x_\mu\delta_{\nu\rho}-x_\rho\delta_{\nu\mu}\Biggr)
+\partial_\nu\Biggl(x_\rho\delta_{\lambda\mu}-x_\mu\delta_{\lambda\rho}
\Biggr)\Biggr]{\cal D}_1^{\rm full}\left(x^2\right), 
\end{equation}
where the average over the monopole densities is defined by the 
partition function~(\ref{zdens}), whereas the $A_\mu$-average 
is defined as 

$$
\left<...\right>_{A_\mu}\equiv\int DA_\mu\left(...\right)
\exp\left(-\frac{1}{4e^2}\int d^3x F_{\mu\nu}^2\right).
$$
In Eq.~(\ref{cora}), 
${\cal F}_{\mu\nu}=F_{\mu\nu}+F_{\mu\nu}^M$ 
stands for the full electromagnetic field strength tensor, which  
includes also the monopole part

$$
F_{\mu\nu}^M\left(\vec x{\,}\right)=-\frac12\varepsilon_{\mu\nu\lambda}
\frac{\partial}{\partial x_\lambda}\int d^3y\frac{\rho\left(\vec y{\,}
\right)}{\left|\vec x-\vec y{\,}\right|}.
$$
This monopole part yields the R.H.S. of 
the Bianchi identities modified by the monopoles, 

\begin{equation}
\label{fullbia}
\partial_\mu {\cal H}_\mu=2\pi\rho,
\end{equation}
where ${\cal H}_\mu=
\frac12
\varepsilon_{\mu\nu\lambda}{\cal F}_{\nu\lambda}$ 
stands for the full magnetic 
induction. Eqs.~(\ref{cora}) and~(\ref{fullbia}) then lead to the 
following equation for the function ${\cal D}^{\rm mon.}$

\begin{equation}
\label{lapld}
\Delta{\cal D}^{\rm mon.}\left(x^2\right)=-4\pi^2\left<\rho
\left(\vec x{\,}\right)\rho(0)\right>_\rho,
\end{equation}
which in fact is a 
3D analogue of the 4D equation~\cite{ufn}

$$
\left(\partial_\mu\partial_\nu-
\partial^2\delta_{\mu\nu}\right){\cal D}^{\rm mon.}\left(x^2\right)=
\left<j_\mu(x)j_\nu(0)\right>.
$$
The correlator standing on the R.H.S. of 
Eq.~(\ref{lapld}) can be found in the limit of small monopole 
densities, $\rho\ll\zeta$. By making use of Eqs.~(\ref{zdens}) 
and~(\ref{lowdens}), we obtain 

$$
\left<\rho
\left(\vec x{\,}\right)\rho(0)\right>_\rho=-\frac{\zeta}{2\pi}
\Delta
\frac{{\rm e}^{-m\left|\vec x{\,}\right|}}{\left|\vec x{\,}\right|}.
$$   
Then, demanding that ${\cal D}^{\rm mon.}\left(
x^2\to\infty\right)\to 0$, we get by the maximum 
principle for the harmonic functions the desired expression for the 
function ${\cal D}^{\rm mon.}$ in the low-density limit 

\begin{equation}
\label{dmon} 
{\cal D}^{\rm mon.}\left(x^2\right)=2\pi\zeta 
\frac{{\rm e}^{-m\left|\vec x{\,}\right|}}{\left|\vec x{\,}\right|}.
\end{equation}
We see that in the model under study, the correlation length of the 
vacuum~\cite{svm, ufn} $T_g$, i.e. the distance at which the function 
${\cal D}^{\rm mon.}$ decreases, 
corresponds to the inverse mass of the field $\chi$, 
$m^{-1}$ ({\it cf.} the case of Abelian-projected theories, studied in 
Refs.~\cite{euro, suzuki}). The coefficient function 
${\cal D}_1^{\rm full}
\left(x^2\right)$ will be derived later on. 

Let us now proceed to the problem of string representation of 3D 
compact QED. To this end, let us 
consider an expression for the Wilson loop 
and try to represent it as an 
integral over the world-sheets $\Sigma$'s, bounded by the contour $C$. 
By virtue of the Stokes theorem, the Wilson loop 
can be rewritten in the form

$$
\left<W(C)\right>=\left<\exp\left(\frac{i}{2}\int
\limits_{\Sigma}^{}d\sigma_{\mu\nu}{\cal F}_{\mu\nu}\right)
\right>_{A_\mu,\rho}=
\left<\exp\left(i\int\limits_{\Sigma}^{}d\sigma_\mu {\cal H}_\mu\right)
\right>_{A_\mu,\rho}=
$$

\begin{equation}
\label{wqed}
=\left<W(C)\right>_{A_\mu}\left<\exp\left(\frac{i}{2}\int d^3x\rho\left(
\vec x{\,}\right)\eta\left(\vec x{\,}\right)\right)\right>_\rho,
\end{equation}
where the free photon contribution reads 

\begin{equation}
\label{wphot}
\left<W(C)\right>_{A_\mu}=
\left<\exp\left(i\oint\limits_{C}^{}A_\mu dx_\mu\right)\right>_{A_\mu}=
\exp\left(-\frac{e^2}{8\pi} 
\oint\limits_{C}^{}dx_\mu\oint
\limits_{C}^{}dy_\mu\frac{1}{\left|\vec x-\vec y{\,}\right|}\right).
\end{equation}
In Eq.~(\ref{wqed}), $d\sigma_\mu\equiv\frac12\varepsilon_{\mu\nu\lambda}
d\sigma_{\nu\lambda}$, and 
$\eta\left(\vec x{\,}\right)=\frac{\partial}{\partial x_\mu}
\int\limits_{\Sigma}^{}d\sigma_\mu\left(\vec y{\,}\right)
\frac{1}{\left|\vec x-\vec y{\,}\right|}$ stands for the 
solid angle under which the surface $\Sigma$ 
is seen by an observer at the point $\vec x$. 
Notice that due to the Gauss law, 
in the case when $\Sigma$ is a closed surface surrounding 
the point $\vec x$, 
$\eta\left(\vec x{\,}\right)=4\pi$, which is the standard result for the 
total solid angle in 3D. 

Eq.~(\ref{wqed}) seems to contain some discrepancy, since its L.H.S. 
depends only on the contour $C$, whereas the R.H.S. depends 
on an arbitrary 
surface $\Sigma$, bounded by $C$. However, this actually occurs to be 
not a discrepancy, but a 
key point in the construction of the desired string representation. 
The resolution of the apparent paradox lies in the observation that 
during the derivation of the effective monopole potential~(\ref{pot}), 
we have accounted only for the one, namely real, branch of the solution 
to the saddle-point equation~(\ref{speq}).   
Actually, however, one should 
sum up over all the (complex-valued) branches of the integrand of the 
effective 
potential~(\ref{pot}) at every space point $\vec x$. This requires 
to replace $V[\rho]$ by 

$$
V_{\rm total}[\rho]=\sum\limits_{\rm branches}^{}
\int d^3x\left\{\pm\rho\ln\left[\frac{\rho}{2\zeta}+
\sqrt{1+\left(\frac{\rho}{2\zeta}\right)^2}\right]\mp 2\zeta
\sqrt{1+\left(\frac{\rho}{2\zeta}\right)^2}
+{0\choose {i\pi}}+2\pi in
\right\},  
$$
$n=0, \pm 1, \pm 2,...$, 
where adding of $0$ or $i\pi$ corresponds to choosing of upper 
or lower sign, respectively. Such a summation over the branches 
of the multivalued potential in the expression for the Wilson 
loop 

$$
\left<W(C)\right>=\left<W(C)\right>_{A_\mu}
\int D\rho
\exp\left\{-\left[\frac{\pi}{2e^2}\int d^3xd^3y
\rho\left(\vec x{\,}\right)\frac{1}{\left|\vec x-\vec y{\,}\right|}
\rho\left(\vec y{\,}\right)+V_{\rm total}[\rho]-
\right.\right.
$$

\begin{equation}
\label{strrepr}
\left.\left.-\frac{i}{2}\int d^3x\rho\left(
\vec x{\,}\right)\eta\left(\vec x{\,}\right)
\right]\right\}
\end{equation} 
thus restores the independence of the choice of the world-sheet. Notice, 
that from now on we omit an inessential normalization factor, 
implying everywhere the normalization $\left<W(0)\right>=1$.

It is worth noting that the obtained string 
representation~(\ref{strrepr}) has been for the first time 
derived in another, more 
indirect, way in Ref.~\cite{confstr}. It is therefore instructive 
to establish a correspondence between our interpretation and 
the one of that paper.

The main idea of Ref.~\cite{confstr} was to calculate 
the Wilson loop starting with the direct definition of this average 
in a sense of the partition function~(\ref{zmon}) of the monopole gas. 
The corresponding expression has the form

$$
\left<W(C)\right>_{\rm mon.}=
$$

$$
=\sum\limits_{N=1}^{+\infty}\sum\limits_{q_a=\pm 1}^{}
\frac{\zeta^N}{N!}\prod\limits_{i=1}^{N}\int d^3z_i
\exp\left[-\frac{\pi}{2e^2}\int d^3xd^3y \rho_{\rm gas}\left(\vec x{\,}
\right)\frac{1}{\left|\vec x-\vec y{\,}\right|}\rho_{\rm gas}\left(
\vec y{\,}\right)+\frac{i}{2}\int d^3x\rho_{\rm gas}
\left(\vec x{\,}\right)\eta\left(\vec x{\,}\right)
\right]=
$$ 

$$
=\int D\chi\exp\left\{-\int d^3x
\left[\frac12\left(\partial_\mu\chi\right)^2-2\zeta
\cos\left(g\chi+\frac{\eta}{2}\right)\right]\right\}=
$$

\begin{equation}
\label{wvarphi}
=\int D\varphi\exp\left\{-\int d^3x\left[\frac{e^2}{8\pi^2}
\left(\partial_\mu\varphi-\frac12\partial_\mu\eta\right)^2-
2\zeta\cos\varphi\right]\right\},
\end{equation}
where $\varphi\equiv g\chi+\frac{\eta}{2}$. 

Next, one can prove 
the following equality

$$
\exp\left[-\frac{e^2}{8\pi}\oint\limits_{C}^{}dx_\mu\oint
\limits_{C}^{}dy_\mu\frac{1}{\left|\vec x-\vec y{\,}\right|}-
\frac{e^2}{8\pi^2}\int d^3x
\left(\partial_\mu\varphi-\frac12\partial_\mu\eta\right)^2\right]= 
$$

\begin{equation}
\label{ramond}
=\int Dh_{\mu\nu}\exp\left[-\int d^3x\left(i\varphi
\varepsilon_{\mu\nu\lambda}\partial_\mu h_{\nu\lambda}+
g^2h_{\mu\nu}^2-2\pi ih_{\mu\nu}\Sigma_{\mu\nu}\right)\right],
\end{equation}
which makes it possible to represent 
the contribution of the 
kinetic term on the R.H.S. of Eq.~(\ref{wvarphi}) and the free 
photon contribution~(\ref{wphot}) 
to the Wilson loop as an integral over the 
Kalb-Ramond field. The only nontrivial point necessary to prove this 
equality is an expression for the 
derivative of the solid angle. 
One has

\begin{equation}
\label{parteta}
\partial_\lambda\eta\left(\vec x{\,}\right)=
\int\limits_{\Sigma}^{}\left(d\sigma_\mu\left(\vec y{\,}\right)
\frac{\partial}{\partial y_\lambda}- 
d\sigma_\lambda\left(\vec y{\,}\right)
\frac{\partial}{\partial y_\mu}\right)\frac{\partial}{\partial y_\mu}
\frac{1}{\left|\vec x-\vec y{\,}\right|}+\int\limits_{\Sigma}^{}
d\sigma_\lambda\left(\vec y{\,}\right)\Delta
\frac{1}{\left|\vec x-\vec y{\,}\right|}.
\end{equation}
Applying to the first integral on the R.H.S. of Eq.~(\ref{parteta}) 
Stokes theorem in the operator form,  

$$
d\sigma_\mu\frac{\partial}{\partial y_\lambda}-
d\sigma_\lambda\frac{\partial}{\partial y_\mu}\longrightarrow 
\varepsilon_{\mu\lambda\nu}dy_\nu, 
$$   
one finally obtains  

$$
\partial_\lambda\eta\left(\vec x{\,}\right)=
\varepsilon_{\lambda\mu\nu}\frac{\partial}{\partial x_\mu}
\oint\limits_{C}^{}dy_\nu\frac{1}{\left|\vec x-\vec y{\,}\right|}-
4\pi\int\limits_{\Sigma}^{}d\sigma_\lambda\left(\vec y{\,}\right)
\delta\left(\vec x-\vec y{\,}\right).
$$ 
Making use of this result and carrying out the Gaussian integral 
over the field $h_{\mu\nu}$, one can 
demonstrate that both sides of Eq.~(\ref{ramond}) are equal to 

$$
\exp\left\{-\frac{e^2}{2}\left[\frac{1}{4\pi^2}\int d^3x
\left(\partial_\mu\varphi\right)^2+\frac{1}{\pi}\int\limits_{\Sigma}^{}
d\sigma_\mu\partial_\mu\varphi+\int\limits_{\Sigma}^{}d\sigma_\mu
\left(\vec x{\,}\right)\int\limits_{\Sigma}^{}d\sigma_\mu
\left(\vec y{\,}\right)\delta\left(\vec x-\vec y{\,}\right)
\right]\right\}
$$
thus proving the validity of this equation. 

Substituting now Eq.~(\ref{ramond}) into Eq.~(\ref{wvarphi}), it 
is easy to carry out the integral over the field $\varphi$, which has 
no more kinetic term, in the saddle-point approximation. This 
equation has the same form as Eq.~(\ref{speq}) 
with the 
replacement $\rho\to\varepsilon_{\mu\nu\lambda}\partial_\mu 
h_{\nu\lambda}$. The resulting expression for the full Wilson loop 
then takes the form

$$
\left<W(C)\right>=\left<W(C)\right>_{A_\mu}
\left<W(C)\right>_{\rm mon.}=
$$

\begin{equation}
\label{wtot}
=\int Dh_{\mu\nu}\exp\left\{-\int d^3x\left(g^2h_{\mu\nu}^2
+V_{\rm total}\left[\varepsilon_{\mu\nu\lambda}
\partial_\mu h_{\nu\lambda}\right]\right)+2\pi i
\int\limits_{\Sigma}^{}d\sigma_{\mu\nu}h_{\mu\nu}\right\},
\end{equation}
where the world-sheet independence of the R.H.S. is again provided 
by the summation over the branches of the multivalued action, which 
is now the action of the Kalb-Ramond field. 

Comparing now Eqs.~(\ref{strrepr}) and~(\ref{wtot}), we see that 
the Kalb-Ramond field is indeed related to the monopole density 
via the equation $\varepsilon_{\mu\nu\lambda}\partial_\mu 
h_{\nu\lambda}=\rho$. Thus, a conclusion following 
from the representation of the 
full Wilson loop in terms of the integral over the Kalb-Ramond field 
is that this field is simply related to the sum of the photon 
and monopole field strength tensors as
$h_{\mu\nu}=\frac{1}{4\pi}
{\cal F}_{\mu\nu}$. In the 
formal language, such a 
decomposition of the Kalb-Ramond field is just the essence of the 
Hodge decomposition theorem.

Let us now consider the weak-field limit of Eq.~(\ref{wtot}) 
and again restrict ourselves to the real branch of the effective 
potential, i.e. replace $V_{\rm total}\left[
\varepsilon_{\mu\nu\lambda}\partial_\mu h_{\nu\lambda}\right]$ by 
$V\left[
\varepsilon_{\mu\nu\lambda}\partial_\mu h_{\nu\lambda}\right]$.
This yields the following expression for the Wilson loop 

\begin{equation}
\label{small}
\left<W(C)\right>_{\rm weak-field}=\int Dh_{\mu\nu}
\exp\left\{-\int d^3x\left[\frac{1}{6\zeta}H_{\mu\nu\lambda}^2+
g^2h_{\mu\nu}^2-2\pi ih_{\mu\nu}\Sigma_{\mu\nu}\right]\right\}.
\end{equation}
Notice, that the mass of the 
Kalb-Ramond field resulting from this equation is equal to the 
mass $m$ of the field $\chi$ from Eq.~(\ref{zcos}).

One can now see that Eq.~(\ref{small})
is quite similar to the 3D version of 
Eq.~(\ref{mon1}) (with the $A_\mu$-field 
gauged away) 
we had in the DAHM case.
However, the important difference from DAHM 
is that restricting ourselves to the 
real branch of the potential, 
we have violated 
the surface independence 
of the R.H.S. of Eq.~(\ref{small}).
This problem is similar to the one which appears in SVM~\cite{svm, ufn}, 
where in the expression for the Wilson loop, written via the non-Abelian 
Stokes theorem and cumulant expansion, one disregards all the 
cumulants higher than the bilocal one (the so-called bilocal 
approximation). There, the surface independence is restored by 
replacing $\Sigma$ by the surface of the minimal area, 
$\Sigma_{\rm min.}=\Sigma_{\rm min.}\left[C\right]$,  
bounded by the contour $C$. 
Let us follow this recipe, after which the quantity 

\begin{equation}
\label{slowen}
\left.S_{\rm str.}=
-\ln\left<W(C)\right>_{\rm weak-field}\right|_{\Sigma\to\Sigma_{\rm min.}}
\end{equation} 
can be considered as a weak-field string 
effective action of 3D compact QED. 

The integration over the Kalb-Ramond field in Eq.~(\ref{small})
is now 
almost the same as the one of Ref.~\cite{euro} and yields 

$$
\left.\left<W(C)\right>_{\rm weak-field}\right|_{\Sigma
\to\Sigma_{\rm min.}}=
\exp\left\{-\frac18
\int\limits_{\Sigma_{\rm min.}}^{}
d\sigma_{\lambda\nu}\left(\vec x{\,}\right)
\int\limits_{\Sigma_{\rm min.}}^{}
d\sigma_{\mu\rho}\left(\vec y{\,}\right)
\left<{\cal F}_{\lambda\nu}\left(\vec x{\,}\right) 
{\cal F}_{\mu\rho}
\left(\vec y{\,}\right)
\right>_{A_\mu,\rho}\right\}, 
$$
which is consistent with the result following directly from the 
cumulant expansion of Eq.~(\ref{wqed}).
Here, the 
bilocal correlator is defined by Eq.~(\ref{cora}) with the 
function ${\cal D}^{\rm mon.}$ given by Eq.~(\ref{dmon}) and 
${\cal D}_1^{\rm full}={\cal D}_1^{\rm phot.}+{\cal D}_1^{\rm mon.}$, 
where the photon and monopole contributions read 

$${\cal D}_1^{\rm phot.}\left(x^2\right)=
\frac{e^2}{2\pi\left|\vec x{\,}\right|^3}$$
and

\begin{equation}
\label{d1mon}
{\cal D}_1^{\rm mon.}\left(x^2\right)=\frac{e^2}{4\pi x^2}
\left(m+\frac{1}{\left|\vec x{\,}\right|}\right){\rm e}^{-m
\left|\vec x{\,}\right|},
\end{equation}
respectively.
Since the 
approximation $\rho\ll\zeta$, in which Eq.~(\ref{dmon}) has been derived, 
is just the weak-field limit, in which 
Eq.~(\ref{small}) follows from Eq.~(\ref{wtot}), 
coincidence of 
the function ${\cal D}^{\rm mon.}$, following from the 
propagator of the Kalb-Ramond field, with the one of 
Eq.~(\ref{dmon})  
confirms the consistency of our calculations. 

Notice that by performing an expansion of the nonlocal string 
effective action~(\ref{slowen}) 
in powers of the derivatives w.r.t. the world-sheet 
coordinates $\xi$, one gets the string tension of the Nambu-Goto term 
and the inverse bare coupling constant of the rigidity term, which 
read

\begin{equation}
\label{sigqed}
\sigma=\pi^2\frac{\sqrt{2\zeta}}{g}~~ 
{\rm and}~~ 
\frac{1}{\alpha_0}=-
\frac{\pi^2}{8\sqrt{2\zeta}g^3},
\end{equation}
respectively. Similarly to the corresponding quantities in the 
Abelian-projected $SU(2)$- and $SU(3)$-gluodynamics, found in 
Refs.~\cite{euro} and~\cite{suzuki}, 
both of them are nonanalytic in $g$, which manifests the 
nonperturbative nature of string representation of all the three 
theories. Notice also, that the negative sign of $\alpha_0$ is 
important for the stability of the string 
world-sheets~\cite{diamant}.

We see that the long- and short distance 
asymptotic behaviours of the functions~(\ref{dmon}) and~(\ref{d1mon}) 
have the same properties as the ones of the corresponding 
functions in QCD within SVM~\cite{digiac}.
Namely, at large distances   
both of the 
functions~(\ref{dmon}) and~(\ref{d1mon}) decrease exponentially 
with the correlation length $m^{-1}$, and at such distances 
${\cal D}_1^{\rm mon.}\ll {\cal D}^{\rm mon.}$ due to the preexponential 
factor. In the same time, in the opposite case $\left|\vec x{\,}
\right|\ll m^{-1}$, the function ${\cal D}_1^{\rm mon.}$ is much 
larger than the function ${\cal D}^{\rm mon.}$, which also  
parallels the SVM results. Notice, however, that the short-distance 
similarity takes place only to the lowest order of perturbation theory 
in QCD, 
where its specific non-Abelian properties are not important. 

It is also worth noting, that the above described 
asymptotic behaviours of the 
functions ${\cal D}^{\rm mon.}$ and ${\cal D}_1^{\rm mon.}$ 
match those of the corresponding functions, which parametrize the 
bilocal correlator of the dual field strength tensors in 
DAHM~\cite{euro}. This similarity, as well as the similarity of 
Eqs.~(\ref{mon1}) and~(\ref{small}), 
tells us that there should 
exist some relation between 3D compact QED and 3D AHM. In what  
follows, we shall demonstrate that such a relation really exists, 
namely 3D compact QED corresponds to the case of small gauge boson 
mass in the London limit of 3D AHM with monopoles. 
Let us stress that in 3D, monopoles are considered as particles at rest,
contrary to the 4D case, where they are generally treated as world-lines 
of moving particles. That is why, in 
order to end up with 3D 
compact QED (i.e. the partition function~(\ref{zmon}) 
of the monopole gas), one should start with 3D AHM with the scalar 
density $\rho_{\rm gas}$ of external 
monopoles at rest, rather than with DAHM. 
The corresponding partition function has the form

\begin{equation}
\label{ahm3}
{\cal Z}_{\rm 3D{\,}AHM}=\int DA_\mu D\theta^{\rm sing.}
D\theta^{\rm reg.}\exp\left\{-\int d^3x\left[\frac{1}{4e^2}
{\cal F}_{\mu\nu}^2+\frac{\eta^2}{2}
\left(\partial_\mu\theta-A_\mu\right)^2\right]\right\}.
\end{equation}
Here, the full field strength tensor again reads 
${\cal F}_{\mu\nu}=F_{\mu\nu}+F_{\mu\nu}^M$, where 
the monopole part 
obeys the equation~(\ref{fullbia}) with the replacement 
$\rho\to\rho_{\rm gas}$. 
Making use of the 
relation~(\ref{vortex}) (where $L$'s are now open lines of 
magnetic vortices,  
ending  
at monopoles and antimonopoles), one can perform the path-integral 
duality transformation of the partition function~(\ref{ahm3}), which 
yields   

$$
{\cal Z}_{\rm 3D{\,}AHM}=\int D\varphi Dy_\mu(\tau) Dh_\mu
\exp\Biggl\{-\int d^3x\Biggl[\frac{1}{4\eta^2}\left(\partial_\mu
h_\nu-\partial_\nu h_\mu\right)^2-2\pi ih_\mu\delta_\mu+
$$

$$
+\left(\frac{e}{\sqrt{2}}h_\mu+\partial_\mu\varphi\right)^2+
\frac{i}{e\sqrt{2}}\left(\frac{e}{\sqrt{2}}h_\mu+
\partial_\mu\varphi\right)\frac{\partial}{\partial x_\mu}
\int d^3y\frac{\rho_{\rm gas}\left(\vec y{\,}\right)}{\left|\vec x-
\vec y{\,}\right|}\Biggr]\Biggr\}
$$
({\it cf.} Eq.~(\ref{threed})). Performing again the gauge 
transformation, which eliminates the field $\varphi$, and integrating 
over the field $h_\mu$, we obtain 

$$
{\cal Z}_{\rm 3D{\,}AHM}=\int Dy_\mu(\tau)\exp\Biggl\{-\frac{\pi\eta^2}{8}
\int d^3xd^3y
\frac{{\rm e}^{-m_A\left|\vec x-\vec y{\,}\right|}}{\left|\vec x-
\vec y{\,}\right|}\Biggl[4\pi\left(\frac{1}{m_A^2}\rho_{\rm gas}
\left(\vec x{\,}
\right)\rho_{\rm gas}
\left(\vec y{\,}\right)+\delta_\mu\left(\vec x{\,}\right)
\delta_\mu\left(\vec y{\,}\right)\right)+
$$

$$
+\int d^3z\frac{\rho_{\rm gas}\left(\vec x{\,}\right)\rho_{\rm gas}
\left(\vec z{\,}
\right)}{\left|\vec y-\vec z{\,}\right|}\Biggr]\Biggr\},
$$
where $m_A=e\eta$ stands for the gauge boson mass. 
The integral 

$$
\int d^3y
\frac{{\rm e}^{-m_A\left|\vec x-\vec y{\,}\right|}}{\left|\vec x-
\vec y{\,}\right|\left|\vec y-\vec z{\,}\right|}=
\int d^3u
\frac{{\rm e}^{-m_A\left|\vec u{\,}\right|}}{\left|
\vec u{\,}\right|\left|\vec x-\vec z-\vec u{\,}\right|}
$$
can easily be calculated by expanding 
$\frac{1}{\left|\vec x-\vec z-\vec u{\,}\right|}$ in Legendre 
polynomials, and the result reads 

$$
\frac{4\pi}{m_A^2\left|\vec x-\vec z{\,}\right|}\left(1-
{\rm e}^{-m_A\left|\vec x-\vec z{\,}\right|}\right).
$$
Taking this into account, we can write down the final expression 
for the partition function of 3D AHM with external 
monopoles, which has the 
following simple form 

\begin{equation}
\label{finahm}
{\cal Z}_{\rm 3D{\,}AHM}=\int Dy_\mu(\tau)\exp\Biggl\{-\frac{\pi\eta^2}{2}
\int d^3xd^3y\Biggl[
\frac{{\rm e}^{-m_A\left|\vec x-\vec y{\,}\right|}}{\left|\vec x-
\vec y{\,}\right|}\delta_\mu\left(\vec x{\,}\right)\delta_\mu
\left(\vec y{\,}\right)+\frac{1}{m_A^2}\frac{\rho_{\rm gas}
\left(\vec x{\,}
\right)\rho_{\rm gas}\left(\vec y{\,}\right)}{\left|\vec x-\vec y{\,}
\right|}\Biggr]\Biggr\}.
\end{equation}
The first term in square brackets on the R.H.S. of Eq.~(\ref{finahm}) 
represents again the 
Biot-Savart interaction between the points of the magnetic 
vortex (and also interaction between vortices, 
if we included several ones), which is 
Yukawa-type, i.e. their Coulomb interaction is screened by the 
condensate of electric Cooper pairs. Contrary, the interaction between 
{\it external} monopoles, represented by the last term on the R.H.S. of 
Eq.~(\ref{finahm}) remains to be unscreened. 

We now see, that when the gauge boson mass becomes small 
(i.e., for example, the magnetic coupling constant $g=\frac{2\pi}{e}$ 
becomes large), 
the Biot-Savart term can be disregarded w.r.t. the interaction of 
external monopoles.
In this limit, 

$$
{\cal Z}_{\rm 3D{\,}AHM}\longrightarrow\exp\left[-\frac{\pi}{2e^2}
\int d^3xd^3y
\frac{\rho_{\rm gas}\left(\vec x{\,}
\right)\rho_{\rm gas}\left(\vec y{\,}\right)}{\left|\vec x-\vec y{\,}
\right|}\right],
$$ 
which is just 
the statistical weight of the partition function of the 
monopole gas~(\ref{zmon}). 
Clearly, this result is in agreement with that of the corresponding 
limiting procedure applied directly to Eq.~(\ref{ahm3}).

\section{Summary}

In the present paper, we have addressed two problems. The first of them 
was the investigation of the 
relation between confinement in the Abelian-projected $SU(2)$- and 
$SU(3)$-gluodynamics and the 
interactions  
between magnetic monopole currents and electric strings. 
To study this problem, 
we have casted the partition function of the 4D Abelian-projected 
$SU(2)$-gluodynamics, which is argued to be just the dual Abelian Higgs 
model, into the form of the integral over the monopole currents. 
Besides the part, quadratic in these currents, 
the resulting monopole effective action turned out to contain also 
a term, which described the interaction of a monopole current with the 
electric ANO string. 
Then, we have extended our analysis 
to the case of the Abelian-projected 4D $SU(3)$-gluodynamics, where 
the resulting representation turned out to contain three 
monopole currents linked to two independent 
string world-sheets in a certain way.
Finally, for illustrations, we have also performed 
the corresponding calculation in 3D, where the role of the moving string 
world-sheets is played by the static electric vortex lines, and the found 
expressions are more transparent. 

The second topic, studied in this paper, was the investigation of 3D 
compact QED and its relation to SVM and 3D Abelian Higgs model with 
external monopoles. Firstly, we have demonstrated that the string 
representation of 3D compact QED (the so-called confining string 
theory) is nothing else, but the integral over the monopole densities. 
Secondly, in the weak-field 
limit of 3D compact QED, we have calculated two coefficient functions, 
which parametrize the bilocal correlator of the field strength tensors 
in the analogous case of SVM. 
One of them has been found by two methods: from the correlator 
of the monopole densities and by virtue of the weak-field limit of the 
confining string theory. Coincidence of both results thus confirms 
the consistency of our calculations. By making use of this   
function, we have then 
obtained the string tension of the Nambu-Goto term 
and the inverse bare coupling constant of the 
rigidity term, corresponding 
to the weak-field effective action of the confining string theory. 
Those turned out to be nonanalytic in the magnetic coupling constant 
(i.e. explicitly nonperturbative) and positive and negative, respectively, 
which is important for the stability of the obtained string effective 
action.  
The large-distance asymptotic 
behaviours of both coefficient 
functions correspond to the ones parametrizing 
the bilocal correlator of the field strength tensors in QCD within SVM. 
The obtained asymptotic behaviours are also similar to those of 
the corresponding functions in DAHM. Finally, we have proved that 
the latter similarity is not accidental, namely 3D compact QED is related 
to the small gauge boson mass limit of 3D AHM with external 
monopoles. 

In conclusion, 
the obtained results shed some light on the mechanisms  
of confinement in various Abelian-projected theories.
Furthermore, they 
prove the relevance of concepts of SVM to the description of 
confinement in 3D compact QED.

\section{Acknowledgments}

We are grateful to H. Kleinert for bringing our attention to 
Refs.~\cite{kleinert} and~\cite{for}. One of us (D.A.) 
would also like to thank the theory group of the Quantum Field 
Theory Department of the Institute of Physics of the Humboldt 
University of Berlin for kind hospitality and Graduiertenkolleg 
{\it Elementarteilchenphysik} for financial support. This work is 
also partly supported by DFG-RFFI, grant 436 RUS 113/309/7.

\section*{Appendix A. Path-Integral Duality Transformation}

In this Appendix, we shall outline some details of a derivation of 
Eq.~(\ref{mon1}). 
Firstly, let us linearize the term $\frac{\eta^2}{2}
\left(\partial_\mu\theta-2gB_\mu
\right)^2$ in the exponent on the R.H.S. of Eq.~(\ref{odin}) 
and carry out the integral over 
$\theta^{{\rm reg.}}$ as follows 

$$\int D\theta^{{\rm reg.}}\exp\left\{-\frac{\eta^2}{2}
\int d^4x \left(\partial_\mu\theta-2gB_\mu
\right)^2\right\}=$$

$$=\int DC_\mu D\theta^{{\rm reg.}}
\exp\left\{\int d^4x\left[-\frac{1}{2\eta^2}C_\mu^2+iC_\mu
\left(\partial_\mu\theta-2gB_\mu
\right)\right]\right\}=$$

$$
=\int DC_\mu\delta\left(\partial_\mu C_\mu\right)
\exp\left\{\int d^4x\left[-\frac{1}{2\eta^2}C_\mu^2+iC_\mu
\left(\partial_\mu\theta^{{\rm sing.}}-2gB_\mu 
\right)\right]\right\}. \eqno (A.1)
$$
Next, one can resolve the constraint 
$\partial_\mu C_\mu=0$ by 
representing $C_\mu$ in the form $C_\mu=\partial_\nu\tilde h_{\mu\nu}
\equiv \frac12
\varepsilon_{\mu\nu\lambda\rho}\partial_\nu h_{\lambda\rho}$, where 
$h_{\lambda\rho}$ stands for an antisymmetric tensor field. 

Notice, that the field $C_\mu$ is related to the 
monopole 
current~(\ref{jmon}) as $C_\mu=-\frac{1}{g}j_\mu$.
This means that the $\delta$-function in the last equality on the 
R.H.S. of Eq.~(A.1) just imposes the conservation of this 
current.

Next, taking 
into account the relation~(\ref{dva}) 
between $\theta^{{\rm sing.}}$ and 
$\Sigma_{\mu\nu}$, we get

$$\int D\theta^{{\rm sing.}}
D\theta^{{\rm reg.}}\exp\left\{-\frac{\eta^2}{2}
\int d^4x \left(\partial_\mu\theta-2gB_\mu
\right)^2\right\}=$$

$$
=\int Dx_\mu (\xi) Dh_{\mu\nu}
\exp\left\{\int d^4x 
\left[-\frac{1}{12\eta^2}H_{\mu\nu\lambda}^2+i\pi h_{\mu\nu}
\Sigma_{\mu\nu}-ig\varepsilon_{\mu\nu\lambda\rho}B_\mu
\partial_\nu h_{\lambda\rho}\right]\right\}. \eqno (A.2)
$$
In the derivation of Eq.~(A.2), 
we have replaced $D\theta^{{\rm sing.}}$ by 
$Dx_\mu(\xi)$ (since the surface $\Sigma$, parametrized by 
$x_\mu(\xi)$, is just the surface, at which the field 
$\theta$ is singular), discarding for simplicity 
the Jacobian~\cite{zubkov} arising during 
such a change of the integration variable.

Bringing together Eqs.~(\ref{odin}) and~(A.2), 
we obtain

$$
{\cal Z}=
\int DB_\mu Dx_\mu(\xi) Dh_{\mu\nu} 
\exp\Biggl\{-\int d^4x\Biggl[\frac1{12\eta^2}H_{\mu\nu
\lambda}^2+\frac14F_{\mu\nu}^2
-i\pi h_{\mu\nu}\Sigma_{\mu\nu}+
ig\tilde F_{\mu\nu}h_{\mu\nu}
\Biggr]\Biggr\}. \eqno (A.3)
$$
Let us now integrate over the field $B_\mu$. 
To this end, we find it convenient 
to rewrite  

$$\exp\left(-\frac14\int d^4x 
F_{\mu\nu}^2\right)=
\int DG_{\mu\nu}\exp\left\{\int d^4x
\left[-G_{\mu\nu}^2+i\tilde F_{\mu\nu}
G_{\mu\nu}\right]\right\},$$
after which the $B_\mu$-integration yields 

$$\int DB_\mu \exp\left\{-\int d^4x\left[\frac14 F_{\mu\nu}^2+
ig\tilde F_{\mu\nu}h_{\mu\nu}\right]\right\}=$$

$$=\int DG_{\mu\nu}\exp\left(-\int d^4x G_{\mu\nu}^2
\right)\delta\left(\varepsilon_{\mu\nu\lambda\rho}\partial_\mu
\left(G_{\lambda\rho}-gh_{\lambda\rho}\right)
\right)=$$

$$
=\int DA_\mu\exp\left[-\int d^4x\left(gh_{\mu\nu}
+\partial_\mu A_\nu-\partial_\nu A_\mu\right)^2\right]. \eqno (A.4)
$$
Here, $A_\mu$ is just the usual gauge field, dual to the 
dual vector potential 
$B_\mu$. Finally, by substituting Eq.~(A.4) into 
Eq.~(A.3), we arrive at Eq.~(\ref{mon1}) of the main text.

\end{document}